\documentclass[aps,pra,10pt,notitlepage,a4paper,superscriptaddress]{revtex4-1}

\usepackage{graphicx}
\usepackage{amsmath}
\usepackage{amsfonts}
\usepackage{color}

\def\<{\langle}
\def\>{\rangle}
\def\({\left(} 
\def\){\right)} 
\def\beq{\begin{equation}}
\def\eeq{\end{equation}}
\def\E#1{10^{#1}}

\usepackage[bookmarksnumbered, bookmarksopenlevel=2, pdfpagelayout=OneColumn,
pdfauthor={Giovanni Ramirez Garcia (giovanni.ramirez@uam.es)}, pdftitle={From
  conformal to volume-law for the entanglement entropy in exponentially
  deformed critical spin 1/2 chains}, pdfsubject={Show that deformations to 1D
  critical Hamiltonians gives rise to volume-law entanglement},
pdfkeywords={entanglement, volume-law, xx model, heisenberg model, critical
  ground states, maximally entangled ground states}]{hyperref}

\begin{document}

\date{\today}

\title{From conformal to volume-law for the entanglement entropy in
  exponentially deformed critical spin 1/2 chains}

\author{Giovanni Ram\'{\i}rez}%
\email{giovanni.ramirez@uam.es}%
\affiliation{Instituto de F\'{\i}sica Te\'orica UAM/CSIC, Madrid, Spain}%
\author{Javier Rodr\'{\i}guez-Laguna}%
\affiliation{Mathematics Dept., Universidad Carlos III de Madrid, Spain}%
\author{Germ\'an Sierra} 
\affiliation{Instituto de F\'{\i}sica Te\'orica UAM/CSIC, Madrid, Spain}%

\begin{abstract}
  An exponential deformation of 1D critical Hamiltonians gives rise to ground
  states whose entanglement entropy satisfies a volume-law.  This effect is
  exemplified in the XX and Heisenberg models.  In the XX case we characterize
  the crossover between the critical and the maximally entangled ground state
  in terms of the entanglement entropy and the entanglement spectrum.
\end{abstract}

\maketitle

\section{Introduction}
\label{sec:introduction}
The ground states (GSs) of local quantum lattice Hamiltonians usually satisfy
an {\em area law} according to which the entanglement entropy $S_A$ of a block
$A$ of the system is proportional to the size of its boundary
\cite{Srednicki.PRL.93, Eisert.RMP.10}.  In one spatial dimension, the area
law means that $S_A$ is bounded by a constant independent on the size of $A$.
This statement was proved by Hasting in 2007, assuming that the Hamiltonian
has finite range (locality), with finite interaction strengths and a gap in
the spectrum \cite{Hastings.JSTAT.07}.  Violations of the area law in 1D
should therefore come from sufficiently non local Hamiltonians, divergent
interaction strengths or gapless systems.  The latter category is the most
studied one and it includes translational invariant critical systems, which
are described by conformal field theory (CFT), for which the gap decays with
the system size $L$ as $1/L$.  In this case a logarithmic violation of the
area law takes place, with a coefficient proportional to the central charge of
the underlying CFT \cite{Wilczek.NPB.94, Vidal.PRL.03,
  Cardy_Calabrese.JSTAT.04}.  The area law is nevertheless restored by a
massive perturbation leading to an entanglement entropy proportional to the
logarithm of the correlation length in the scaling regime
\cite{Cardy_Calabrese.JSTAT.04}.

In this work we investigate a much stronger violation of the area law.  We
deform 1D critical Hamiltonians with open boundary conditions (OBC), choosing
couplings that decay exponentially outwards of both sides of the center of the
chain \cite{Vitagliano.NJP.10}.  This decrease of the couplings yields a
vanishing gap in the thermodynamic limit, allowing for a violation of the area
law, that turns into a volume law.  If the decay of the couplings is very
fast, one can use the Dasgupta-Ma renormalization group (RG) that has been
applied successfully to strong disordered systems \cite{Dasgupta_Ma.80}.  The
resulting GS turns out to be a valence bond state formed by bonds joining the
sites located symmetrically with respect to the center.  This state was termed
{\em concentric singlet phase} by Vitagliano et al. \cite{Vitagliano.NJP.10},
and has a {\em rainbow}-like structure as illustrated in figure
\ref{fig:rb.scheme}.

The role of coupling inhomogeneity in 1D quantum many-body physics has been
addressed from many different points of view.  Quenched disorder in the
couplings gives rise to GSs, which, when averaged, resemble quantum critical
states \cite{Refael_Moore.PRL.04, Refael_Moore.JPA.09, Laflorencie.PRB.05,
  Fagotti.PRB.11, Ramirez.JSTAT.14}.  If the couplings change smoothly enough,
they can be regarded as a position-dependent speed of propagation for the
excitations, or a local gravitational potential \cite{Boada.NJP.11}.  Thus, a
slow decrease of the couplings to zero can be regarded as a {\em horizon}
\cite{Laguna.14}.  Smoothed boundary conditions, in which the couplings fall
to zero in the borders, have been used to reduce the finite-size effects when
measuring bulk properties of the GS \cite{Vekic.PRL.93}.  The opposite case,
in which the couplings increase exponentially or hyperbolically, has also been
studied in the literature \cite{Okunishi.PRB.10, Nishino.JPSJ.09,
  Nishino.PTP.10}.  Other possible violations of the area-law have been
recently investigated in reference \cite{Trombettoni.14} by means of
long-range couplings with a magnetic phase and a Fermi surface with a point of
accumulation.

The aim of this work is to study the behavior of entanglement in exponentially
deformed critical 1D models: XX and Heisenberg.  Concretely, we will focus on
the transition from the conformal to the volume law.

This article is organized as follows.  In section \ref{sec:rainbow}, we define
the deformed XX model in the free fermion formulation, apply the Dasgupta-Ma
RG method to construct the GS for a strong deformation and obtain the
numerically exact solution for all the deformation strengths.  In section
\ref{sec:entanglement}, we compute the entanglement entropy and analyze its
scaling properties, inspired by CFT and the transition between the conformal
to the rainbow state.  In section \ref{sec:spectrum}, we analyze the
entanglement spectrum and the deformed Heisenberg model, obtaining similar
results.  Finally, in section \ref{sec:conclusions} we present the conclusions
and points to further work.

\section{The model}
\label{sec:rainbow}
\begin{figure}
  \centering
  \begin{minipage}[c]{80mm}
    \includegraphics[width=80mm]{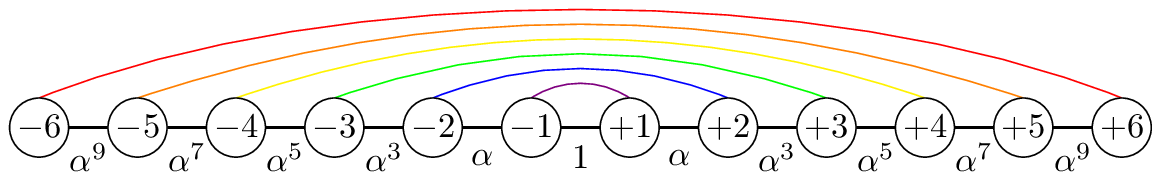}
  \end{minipage}%
  \hspace{10mm}%
  \begin{minipage}[c]{40mm}
    \includegraphics[width=40mm]{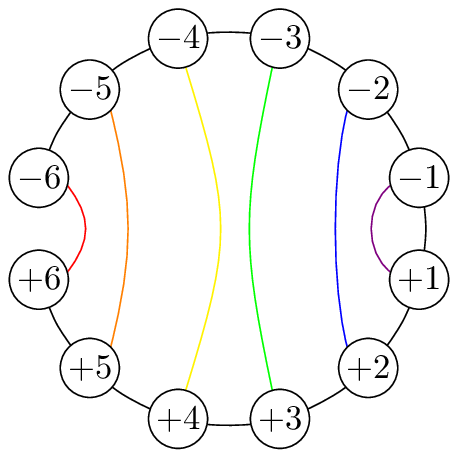}
  \end{minipage}
  \caption{(Color online) Rainbow state both in linear and circular
    representations, showing the $(-k,+k)$ bonds above the central link.
    Thus, the entanglement entropy of the left (or right) half of the chain is
    $L \log 2$.}
  \label{fig:rb.scheme}
\end{figure}

Let us consider a chain with $2 L$ sites labelled by integers $i= \pm 1,
\dots, \pm L$.  The Hamiltonian of the model is given by
\begin{equation}
  H_L \equiv -  J_0 c_1^\dagger  c_{-1} -  \sum_{i=1}^{L-1} J_i  \left(
    c^\dagger_i c_{i+1} + c^\dagger_{-i}  c_{-(i+1)}  \right)  + \mathrm{h.c.}
  \label{eq:rb.hamiltonian}
\end{equation}
where $c_i$ and $c^\dagger_i$ are annihilation and creation operators of
spinless fermions and $J_i$ are the hopping amplitudes parametrized as (see
figure \ref{fig:rb.scheme})
\begin{equation}
  \begin{cases}
    J_0(\alpha)=1,  &  \\ 
    J_i(\alpha)= \alpha^{2 i-1}, & i=1,  \dots,  L-1
  \end{cases}  
  \label{eq:rb.couplings}
\end{equation}

Via a Jordan-Wigner transformation, the Hamiltonian \eqref{eq:rb.hamiltonian}
is equivalent to the XX model for a spin 1/2 chain.  For $\alpha =1$, one
recovers the well known uniform spinless fermion model with OBC.  The model
with $0 < \alpha < 1$ was introduced by Vitagliano et al. to illustrate a
violation of the area law for a local Hamiltonian \cite{Vitagliano.NJP.10}.
Taking $\alpha >1$ and truncating the chain to the sites $i=1, \dots, L$, one
obtains the Hamiltonian considered by Okunishi and Nishino, which has the
scale-free property of the Wilson's numerical renormalization group of the
Kondo impurity problem \cite{Okunishi.PRB.10}.  The models where $J_i$ is a
hyperbolic function have been considered in order to measure the energy gap
\cite{Nishino.PTP.10}.

The ground state of $H_L$ can be studied by means of two different methods:
Dasgupta-Ma RG and exact diagonalization.  The former provides a valence bond
picture of the GS in the limit $\alpha \rightarrow 0^+$, wich explains in
simple terms the volume law.  On the other hand, the second method is
applicable to all values of $\alpha$ and in particular to the limit $\alpha
\rightarrow 1^-$, where one recovers the uniform model, with a log law
described by CFT.  We shall show that the two limits are connected
continuously, that is, with no phase transitions between them.  This fact
offers the possibility of studying the crossover between the log law and the
volume law of the entanglement entropies, which exhibits interesting features.

{\em 1.- Renormalization Group.} In the Dasgupta-Ma RG method, one selects the
strongest link between two nearest neighbor sites and places a bonding state
between them \cite{Dasgupta_Ma.80}
\begin{equation}
 |\Psi \rangle = \frac{1}{\sqrt{2}} \( |10 \rangle +  |01 \rangle  \).
  \label{eq:singlet}
\end{equation}

The two sites are then removed from the chain and an effective coupling is
established between the two sites that border the newly created bond.  The
hopping amplitude for this effective link is found using second order
perturbation theory:
\begin{equation}
  \tilde J_i = \frac{J_{i-1} J_{i+1}}{J_i}. 
  \label{eq:dasgupta-ma}
\end{equation}

In the framework of tensor networks, this renormalization step can be seen as
a disentangler operation between the sites contained in the bond
\cite{Cirac_Verstraete.JPA.09}.  In our case, that is $\alpha < 1$, the
strongest link is the central one.  A bond is thus established on top of link
$J_0$, between sites $-1$ and $+1$ and an effective hopping is established
between sites $-2$ and $+2$, with strength
\begin{equation}
  \tilde J_{-2,+2} = \frac{J_{+1}J_{-1}}{J_{0}} = \alpha^2.
  \label{eq:effective.bond}
\end{equation}

This newly created link is again the strongest one, since $\alpha^2>\alpha^3$,
so we establish a new bond between sites $-2$ and $+2$ (figure
\ref{fig:rb.scheme}, blue bond), and renormalize the coupling between sites
$-3$ and $+3$ as $\alpha^4$.  The procedure repeats itself, until the $2L$
sites are linked by $L$ valence bond states, whose shape looks like a {\em
  rainbow}.  It is easy to see that after the first RG state one can factor
out an overall constant $\alpha^2$ in the couplings $J_i$, such that the
renormalized Hamiltonian becomes $\alpha^2 H_{L-1} + {\rm const}$.  This fact
implies that the rainbow state is a trivial fixed point of the RG with zero
correlation length between nearest neighbor sites, except the sites $i= \pm
1$.

In summary, the rainbow state can be written as the valence bond state:
\begin{equation}
 |R_L \rangle  = \prod_{k=1}^L |\Psi \rangle _{-k,+k}\,, 
  \label{eq:rb.state}
\end{equation}
where $|\Psi \rangle _{-k,+k}\,$, given in equation \eqref{eq:singlet}, is a
single bond, a {\em Bell pair} between sites $-k$ and $+k$.  In that state,
the reduced density matrix $\rho_B$ of any block $B$ has a very characteristic
spectrum $\{\lambda_p\}$ \cite{Refael_Moore.JPA.09}: if $n_B$ is the number of
bonds joining $B$ with the rest of the system, the eigenvalue $2^{-n_B}$
appears with multiplicity $2^{n_B}$.  Thus, the von Neumann entropy can be
easily computed: $S(B)\equiv -\sum \lambda_k \log \lambda_k =n_B\log 2$, i.e.
the number of broken Bell pairs multiplied by $\log 2$.  Moreover, all R\'enyi
entropies take the same value.  Within the RG approximation, the entanglement
properties of the GS are independent of $\alpha$.  The validity of the
renormalization scheme improves when the renormalized link is much stronger
than the surrounding ones, that is $\alpha \ll 1$.  Thus, one can assert that
the rainbow state becomes the exact GS of the $H_L$ Hamiltonian in the limit
$\alpha\to 0^+$.

Let $B$ be the block containing half of the chain, so $n_B=L$.  Its
entanglement entropy is straightforward to compute: $L\cdot \log 2$; i.e. the
state is maximally entangled and fulfills a volume law.  The energy gap can be
estimated as the effective energy of the last bond established, which scales
as $\alpha^{2L}$ and for $\alpha<1$ vanishes in the limit $L \rightarrow
\infty$ in agreement with the Hastings theorem.  For use later, we shall
define the following quantity
\beq
z = - L \log \alpha, 
\label{eq:z}
\eeq 
in terms of which $\alpha^{2L}= e^{-2z}$.  Notice that $z$ can be endowed with
a physical interpretation as the ratio between $L$ and the decay length of the
hopping amplitudes.  We shall see below that $z$ plays the role of a scaling
parameter in the limit $L \gg 1$ and $\alpha \approx 1$.

\begin{figure}
  \centering
  \begin{minipage}[c]{80mm}
    \includegraphics[width=80mm]{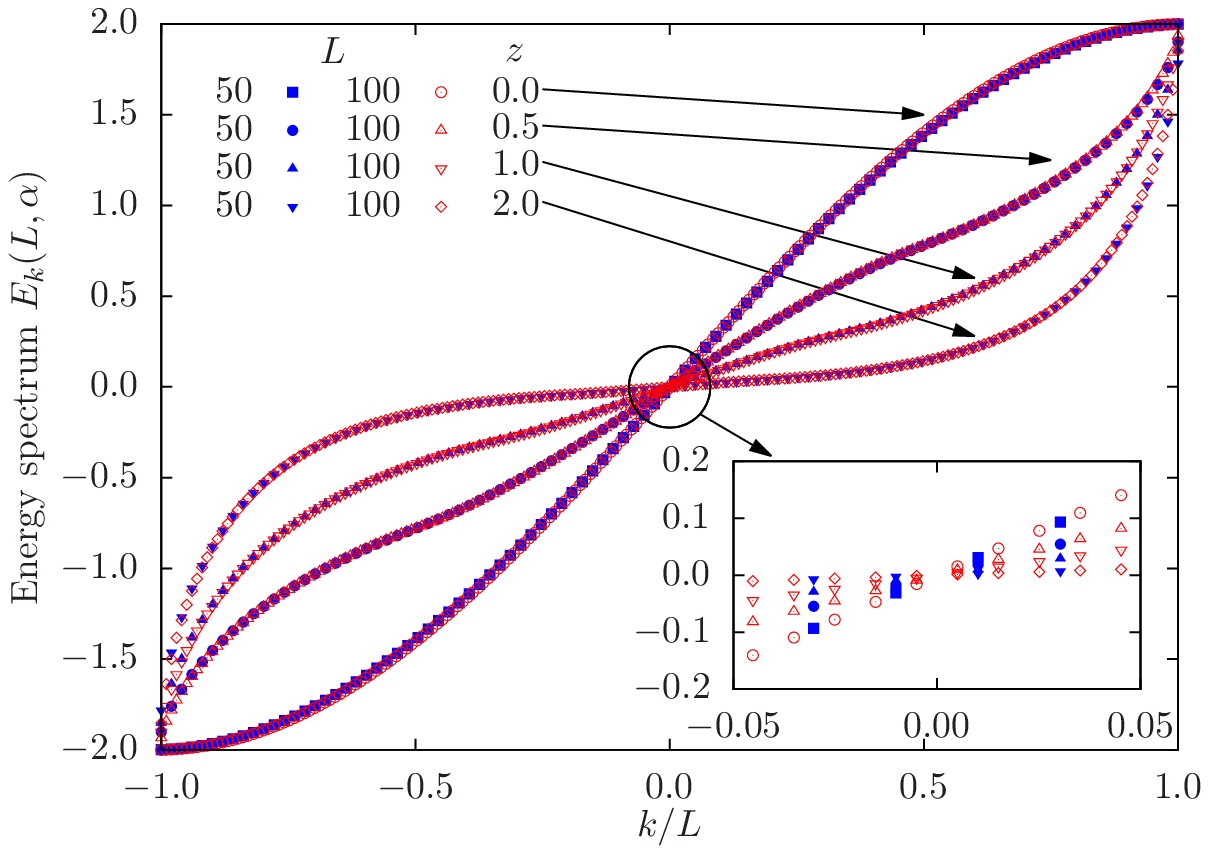}
  \end{minipage}%
  \hspace{10mm}%
  \begin{minipage}[c]{80mm}
    \includegraphics[width=80mm]{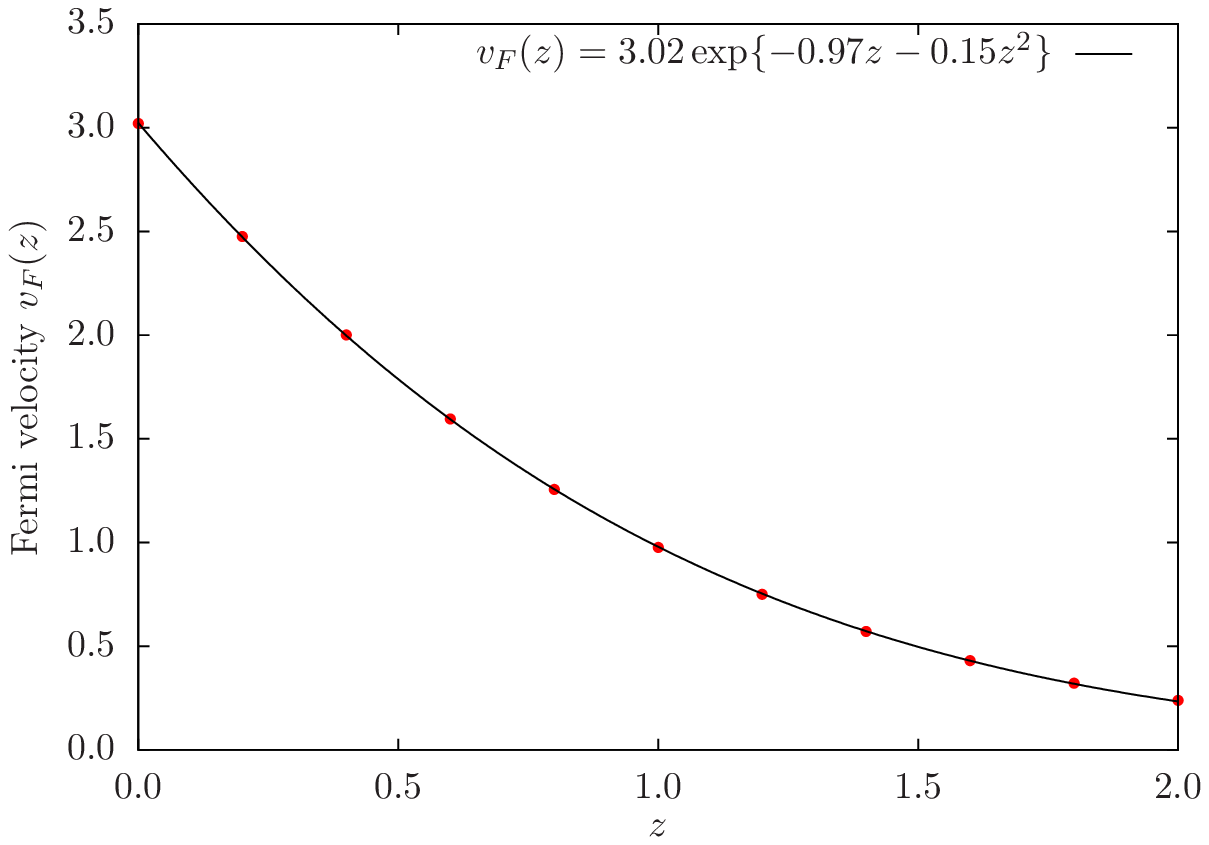}
  \end{minipage}
  \caption{Left: Energy spectrum $E_k(L, \alpha)$ for $L=50, 100$ and several
    values of $z$.  The data collapse on the same curve, which shows the
    scaling law \eqref{eq:scl}.  Right: Fermi velocity $v_F(z)$ as a function
    of $z$.  The solid line is an exponential fit.}
  \label{fig:ek}
\end{figure}

{\em 2.- Exact Diagonalization.}  The Hamiltonian \eqref{eq:rb.hamiltonian} is
quadratic in the fermionic operators.  Therefore its spectrum can be obtained
by diagonalizing the corresponding $2L\times 2L$ hopping matrix $T_{ij}=- J_0
\delta_{i j , -1} -J_i \delta_{|i-j|, 1}, \; i,j =\pm 1, \dots, \pm L$.  One
can easily verify that if $\phi_i$ is an eigenfunction with energy $E$, then
$(-1)^i {\rm sign }(i) \phi_i$ is another eigenfunction with energy $-E$.
Thus the GS of the chain is obtained by filling the lowest energy levels with
$L$ fermions (half-filling).  The eigenmodes, $\phi^k_i$, fulfill $T_{ij}
\phi^k_j = E_k \phi^k_i$.  We shall choose the label of the eigenfunctions as
$k =0, \pm 1, \dots, \pm (L-1), - L$, such that the particle-hole symmetry
becomes $E_{k} = - E_{-k-1}$ and $E_{k} < E_{k+1}$.  In the uniform case, i.e.
$\alpha =1$, one obtains $E_k = 2 \sin \left[ \pi ( 2 k +1)/( 2 (2 L +1))
\right]$.  The GS of the chain is given by
\begin{equation}
 |R(\alpha) \rangle  = \prod_{k=-1}^{-L}  d^\dagger_k(\alpha) |0  \rangle, 
   \label{eq:fermi.sea}
\end{equation}
where $d^\dagger_k = \sum_i \phi^k_i c^\dagger_i$.  We have not found closed
analytic expressions for the eigenvalues and eigenfunctions when $\alpha \neq
1$, but some general properties can be obtained numerically.  In particular,
the spectrum $E_k(L, \alpha)$ satisfies the scaling law
\beq 
  E_k(L, \alpha) \simeq e_z(k/L), \quad k, L \gg 1,
  \label{eq:scl}
\eeq
which is illustrated in the left panel of figure \ref{fig:ek}.  The dispersion
relation $e_z$ changes smoothly from the $sine$ function, for $z \ll 1$ to an
almost flat function near the origin for $z \gg 1$.  At half-filling the
relevant modes lie in the neighborhood of Fermi point where the dispersion
relation linearizes,
\beq
  e_z(k/L) \simeq v_F(z) k/L, \qquad k/L \ll 1, 
  \label{eq:ez}
\eeq
with a Fermi velocity $v_F(z)$ (see right panel of figure \ref{fig:ek}).

\section{Entanglement entropy} 
\label{sec:entanglement}
The entanglement properties are found from equation \eqref{eq:fermi.sea}
\cite{Peschel.JPA.03}.  Let $B$ be a block of size $\ell$, and let $i$, $j\in
B$.  Then the two point-correlator of the fermion operators $c_i$ is given by
\begin{equation}
  C^B_{ij} = \langle R(\alpha)| c^\dagger_i c_j |R(\alpha) \rangle  =
  \sum_{k=-1}^{-L}  \bar\phi^k_i(\alpha) \phi^k_j(\alpha).
  \label{eq:correlation.matrix}
\end{equation}

\begin{figure}
  \centering
  \includegraphics[width=80mm]{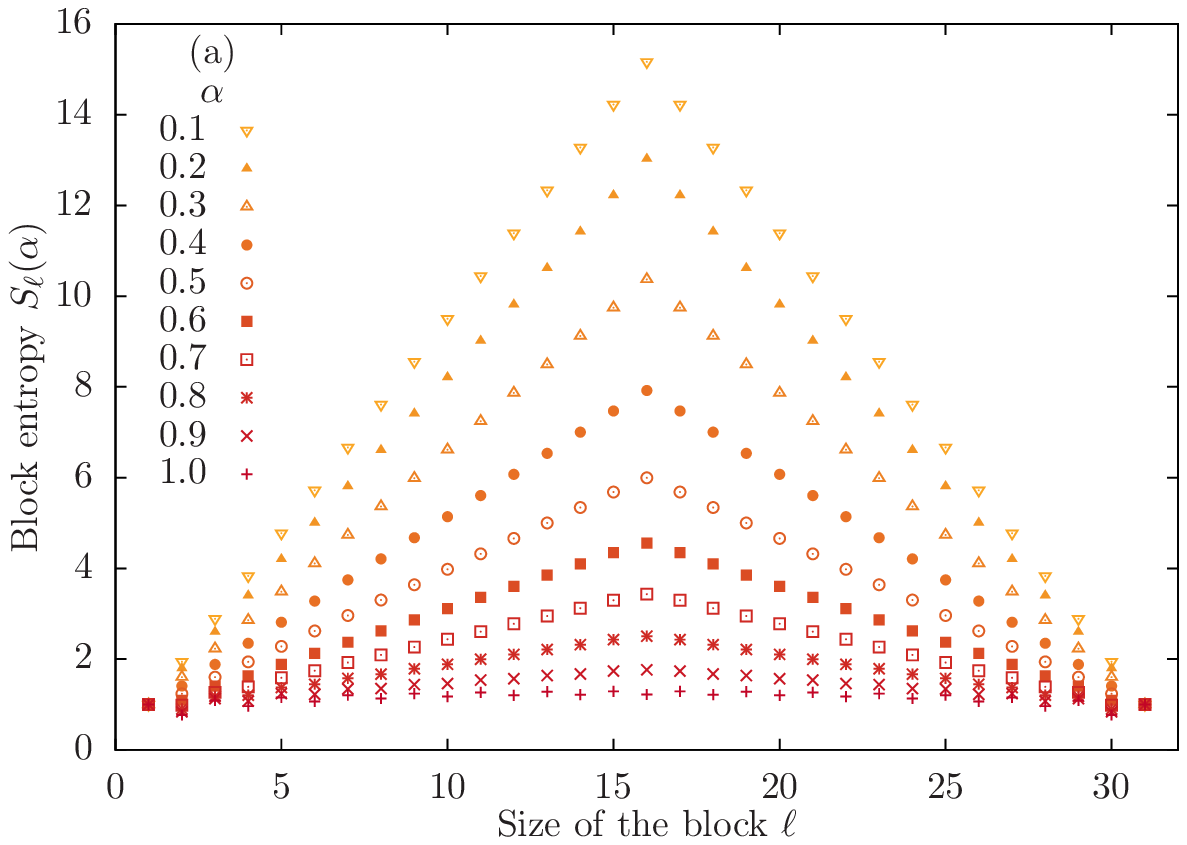}\\\vspace{5mm}
  \begin{minipage}[c]{40mm}
    \includegraphics[width=40mm]{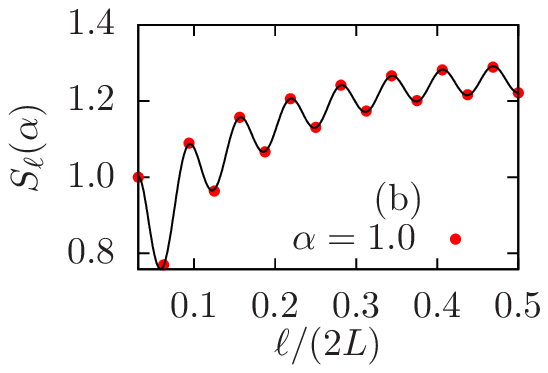}
  \end{minipage}\hspace{2mm}%
  \begin{minipage}[c]{40mm}
    \includegraphics[width=40mm]{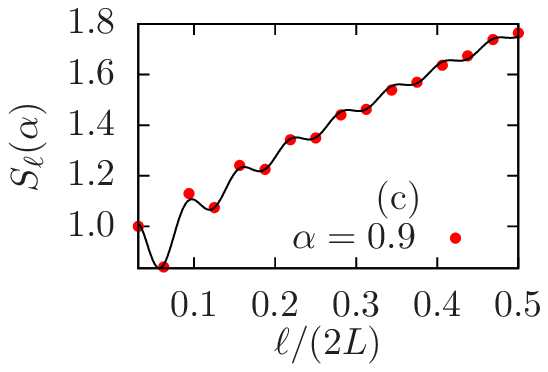}
  \end{minipage}\hspace{2mm}%
  \begin{minipage}[c]{40mm}
    \includegraphics[width=40mm]{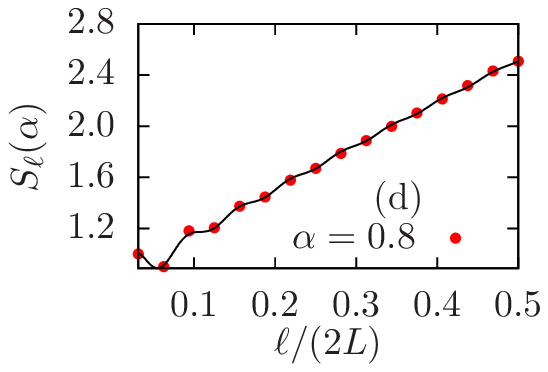}
  \end{minipage}\hspace{2mm}%
  \begin{minipage}[c]{40mm}
    \includegraphics[width=40mm]{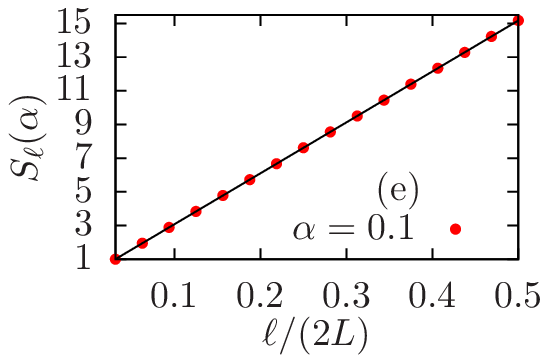}
  \end{minipage}
  \caption{(a) Block entropy $S_\ell(\alpha)$, for a system of size $L=16$
    ($32$ sites).  Notice the {\em tent shape} for small $\alpha$, denoting
    volumetric growth of the entanglement entropy.  (b) For the uniform case
    $\alpha=1$; (c) for $\alpha=0.9$; (d) for $\alpha=0.8$ and (e) for
    $\alpha=0.1$.}
  \label{fig:blockEntropy}
\end{figure}

Let $\{\nu_p\}_{p=1}^\ell$ be its eigenvalues, then the reduced density matrix
of the block can be written as $\rho_B=\otimes_{p=1}^\ell \rho_p$, where
\begin{equation}
  \rho_p=\nu_p b^\dagger_p b_p + (1-\nu_p) b_p b^\dagger_p, 
  \label{eq:entanglement.occ}
\end{equation}
for some fermionic operators $b^\dagger_p$.  The values $\nu_p=\<b^\dagger_p
b_p\>$ are interpreted as {\em occupations} of the different modes
$b^\dagger_p$.  The von Neumann entropy of $B$ is given by
\begin{equation}
  S(B)=-\sum_{p=1}^\ell  \left[ \nu_p \log \nu_p + (1-\nu_p) \log(1-\nu_p)
  \right].
  \label{eq:von_neumann}
\end{equation}

The set $\{\nu_p\}$ allows a full computation of the entanglement spectrum,
i.e. the spectrum of the reduced density matrix $\rho_B$, which provides the
most complete information about entanglement and can help characterize quantum
phase transitions\cite{Li_Haldane.PRL.08}.

We should remark that the numerical computation of the eigenstates of matrix
$T_{ij}$ is an ill-conditioned problem if $z$ is large.  Working at double
precision the upper bound for $z$ can be estimated as $e^{ - 2 z_{\rm max}}
\sim \E{-16}$, that is $z_{max} \sim 18$, but we shall be usually working
below this value.

Let $S_\ell(\alpha)$ denote the von Neumann entropy of the block containing
the leftmost $\ell$ sites in the GS.  Figure \ref{fig:blockEntropy}(a) shows
its dependence with $\ell$ for different values of $\alpha$ in a system with
$L=16$, i.e. with $32$ sites.  For low values of $\alpha$ we observe a
characteristic {\em tent shape}, i.e. an approximately linear growth up to
$\ell=L$ followed by a symmetric linear decrease, giving the volumetric
behavior.  As $\alpha$ grows, the slope decreases and ripples start to appear.
These numerical data can be fitted to a formula that contains linear,
oscillating and logarithmic functions of $\ell$ with coefficients that depend
in a non trivial manner in $\alpha$ and $L$.

\begin{figure}
  \centering
  \includegraphics[width=80mm]{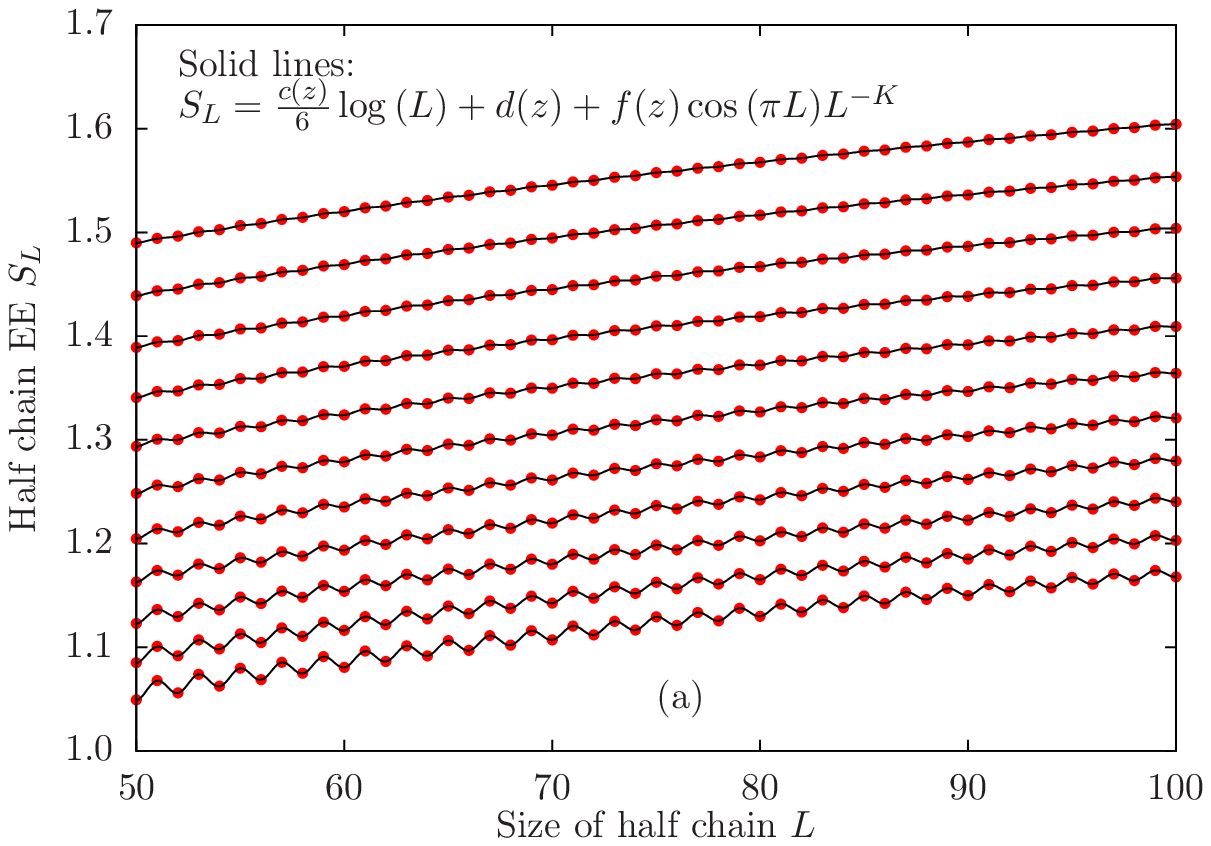}\\\vspace{5mm}
  \begin{minipage}[c]{53mm}
    \includegraphics[width=53mm]{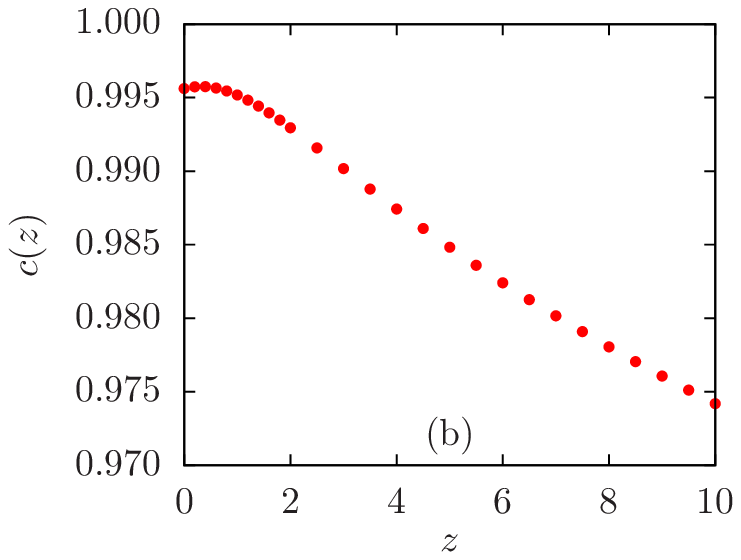}
  \end{minipage}\hspace{3mm}%
  \begin{minipage}[c]{53mm}
    \includegraphics[width=53mm]{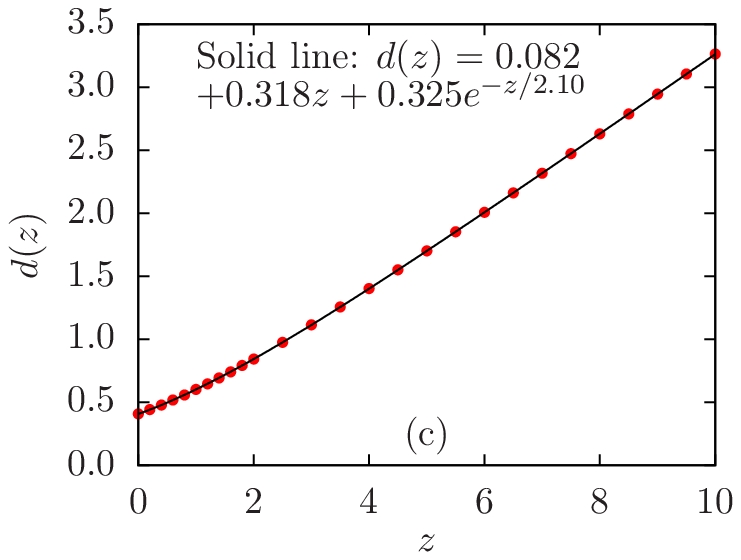}
  \end{minipage}\hspace{3mm}%
  \begin{minipage}[c]{53mm}
    \includegraphics[width=53mm]{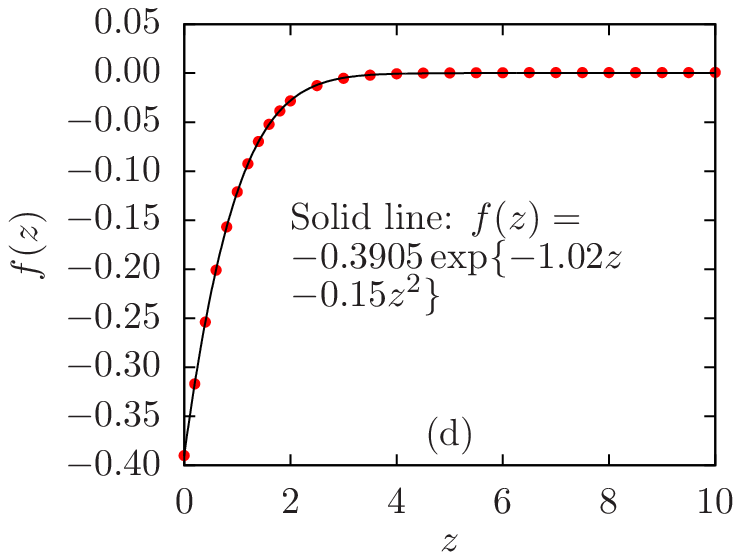}
  \end{minipage}
  \caption{(a) Half-chain entanglement entropy $S_L$ as a function of the
    system half size $L=50, \dots, 100 $ for $z=0$ to $2$ in steps of $0.2$
    from bottom to top.  Solid lines are fits to \eqref{eq:sl} with $\chi^2
    \sim \E{-10}$.  The data corresponding to the values $2 < z \leq 10$, not
    shown in this figure, also satisfy equation \eqref{eq:sl}.  (b)-(d)
    Functions $c(z)$, $d(z)$ and $f(z)$, in the interval $z \in [0, 10]$,
    together with fits for $d(z)$ and $f(z)$.}
  \label{fig:sl}
\end{figure}

To simplify the analysis of the functional dependence of the entropy on these
parameters, we shall consider the von Neumann entropy of the half-chain,
$S_L(\alpha)$.  Figure \ref{fig:sl}(a) shows the values of $S_L$ for $L = 50,
\dots, 100$ and fixed values of $z=0, \dots, 2$ (note that $\alpha$ is tuned
with $L$ in order to keep $z$ constant).  Quite remarkably the half-chain
entropy can be fitted to the expression
\beq
  S_L = \frac{c(z)}{6} \log L + d(z) + f(z) \cos(\pi L) L^{- K} , 
  \label{eq:sl}
\eeq 
where the functions $c(z)$, $d(z)$ and $f(z)$ are shown in figures
\ref{fig:sl}(b)-(d) respectively, together with the corresponding fits.  The
Luttinger parameter $K$ in equation \eqref{eq:sl} is taken equal to $1$, which
gives the best fit to the numerical data.  Equation \eqref{eq:sl} is motivated
by the standard CFT formulas recovered in the case $z=0$, which corresponds to
a CFT with central charge $c=1$ and Luttinger parameter $K=1$
\cite{Vidal.PRL.03, Cardy_Calabrese.JSTAT.04, Laflorencie_etal.PRL.06,
  Calabrese_etal.PRL.10, Fagotti_Calabrese.JSTAT.11}.

Indeed, in the limit $z \rightarrow 0$, we obtain $c(z) \rightarrow 0.995$.
As $z$ increases, the function $c(z)$ decreases.  This result reminds us of
the Zamolodchikov $c-$theorem, according to which a certain function $C$ of
the coupling constants of a relativistic $1+1$ quantum field theory, never
increases along the RG flow and equals the central charge of the CFT at the
fixed points \cite{Zamolodchikov.86, Cardy-book}.  In our case, there is a
fixed point at $z=0$, which corresponds to a free fermion with OBC which has
$c=1$.  One should expect that along the RG flow the value of $z$ increases
while $c(z)$ decreases, approaching zero in the limit $z \rightarrow \infty$,
where one finds the rainbow state, which is a trivial fixed point of the RG.

Let us next discuss the term $d(z)$ in equation \eqref{eq:sl}.  In a CFT on a
strip of length $2L$, the entanglement entropy of the half line is given by
$S_L = c/6 \log (2 L/\pi) + c'_1 + 2 g + f \cos(\pi L) L^{-K}$
\cite{Cardy_Calabrese.JSTAT.04, Laflorencie_etal.PRL.06,
  Calabrese_etal.PRL.10, Fagotti_Calabrese.JSTAT.11}, where $g$ is the
boundary entropy of Affleck and Ludwig \cite{Affleck_Ludwig.91}.  We may then
interpret $d(z)$ as a $z$ dependent boundary entropy $g(z)$, up to some non
universal constants.  As the $c$-theorem, the $g-$theorem asserts that the $g$
function decreases under the RG flow of the boundary, so long as the bulk
theory remains critical during the boundary flow \cite{Affleck_Ludwig.93,
  Friedan_Konechny.04}.  However, there are no reasons for this behavior if
the bulk theory also flows with the RG \cite{GreenMulliganStarr.08}.  This is
the situation found here, where $d(z)$ increases with $z$, as shown in figure
\ref{fig:sl}(c).  The linear increase of $d(z)$ is responsible for the
extensive behavior of the entanglement entropy and can be understood from the
Dasgupta-Ma RG in the large $z$ regime, or $\alpha \ll 1$.

The last term in equation \eqref{eq:sl} describes the oscillations of $S_L$,
which are clearly visible in figure \ref{fig:sl}(a) for $z \leq 1$.  This
behavior is due to the function $f(z)$, which vanishes for $z \simeq 2$ as
shown in figure \ref{fig:sl}(d).

We can use equation \eqref{eq:sl} to study the limit $L \gg 1$ with $\alpha$
kept constant, which implies $z \gg 1$.  From figures \ref{fig:sl}(b)-(d) one
finds that $c(z) \rightarrow 0$, $d(z) \rightarrow 0.318 z$ and $f(z)
\rightarrow 0$ so that
\beq
  S_L \rightarrow  - 0.318  L \log \alpha, \qquad L \gg 1.
  \label{eq:sl2}
\eeq 

This result cannot be valid for very small $\alpha$ since we know that for
$\alpha \to 0^+$ the entropy is given by $S_L = L \log 2$.  The crossover
takes place for $\alpha \sim 1/8$.

\section{Entanglement Spectrum}
\label{sec:spectrum}
In order to provide a thorough characterization of the entanglement of the
half-chain we have analyzed its entanglement spectrum (ES)
\cite{Li_Haldane.PRL.08}.  The reduced density matrix for a block can always
be written as $\rho_B \equiv \exp(-H_E)$, where $H_E$ is called the
entanglement Hamiltonian.  In the case where the state is a Slater
determinant, such as $|R_L(\alpha) \rangle$, $H_E$ can be expressed as a
free-fermion Hamiltonian:
\begin{equation}
  H_E = \sum_{p=1}^\ell \epsilon_p b^\dagger_p b_p + f_0, 
  \label{eq:entanglement_ham}
\end{equation}
where $\epsilon_p$ are the entanglement energies (EE), which can be computed
from the eigenvalues $\nu_p = \<b^\dagger_p b_p\>$ of the correlation matrix
$C^B_{ij}$ as
\begin{equation}
  \nu_p = \frac{1}{1+\exp(\epsilon_p)}.
  \label{eq:occupations}
\end{equation}

\begin{figure}
  \centering
  \begin{minipage}[c]{80mm}
    \includegraphics[width=80mm]{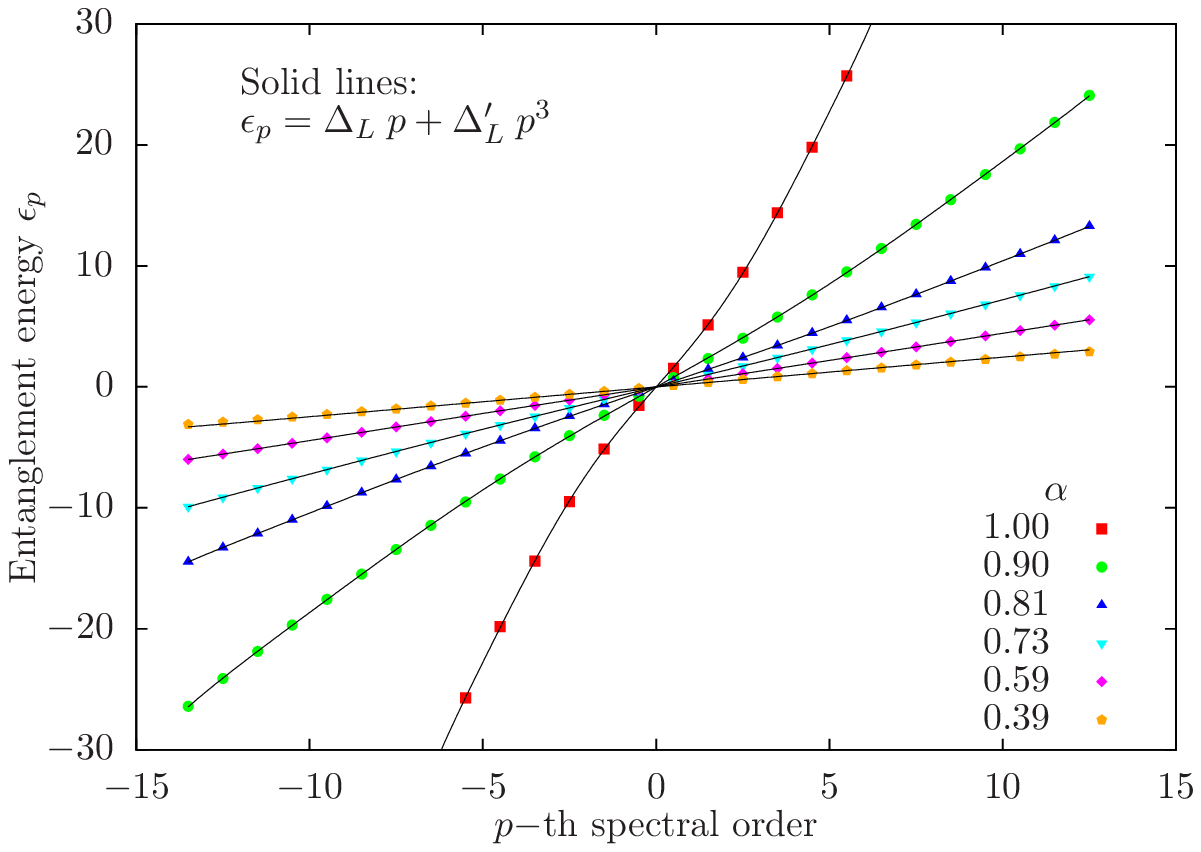}
  \end{minipage}%
  \hspace{10mm}%
  \begin{minipage}[c]{80mm}
    \includegraphics[width=80mm]{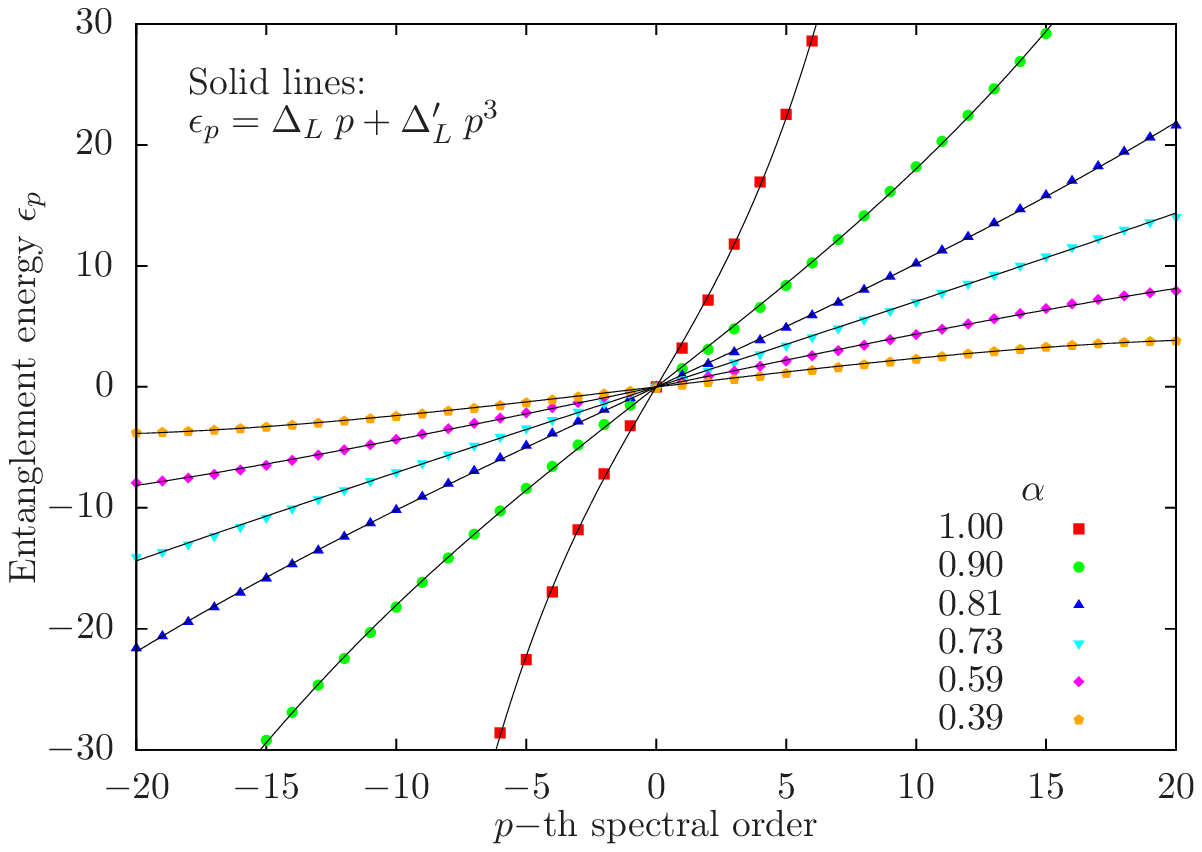}
  \end{minipage}
  \caption{Entanglement energies of the half-chain for several values of
    $\alpha$ and $L=40$ (left) and $L=41$ (right) together with a fit to the
    equation \eqref{eq:es_fit}.}
  \label{fig:EEspectra}
\end{figure}

The overall constant $f_0$ is given by 
\beq
  f_0 = \sum_{p=1}^\ell \log( 1 + e^{\epsilon_p}).
  \label{eq:epsi}
\eeq

Let us consider our block to be the half-chain.  In the limit $\alpha\to 0^+$,
we obtain the rainbow state, which is maximally entangled and the ES is
straightforward to describe.  Each site makes up a bond with another site out
of the block.  Thus, each broken bond provides an entanglement mode,
$b^\dagger_p$, localized at site $p$, with occupation probability $\nu_p=1/2$.
Applying expression \eqref{eq:occupations}, we can see that the entanglement
energies are all $\epsilon_p=0$.  In other terms, the entanglement Hamiltonian
$H_E=f_0 = L \log 2$ gives the entanglement entropy $S_L = L \log 2$.

Figure \ref{fig:EEspectra} shows the ES for a chain with $L=40$ and $L=41$,
for different values of $\alpha$.  Note that for $L$ odd there is a zero
energy.  In agreement with the previous discussion for small $\alpha$, the
values of $\epsilon_p$ are located around zero.  However, as $\alpha$
increases the EE increases almost linearly in the proximity of the zero energy
following the law
\begin{equation}
  \epsilon_p \approx \Delta_L\; p + \Delta'_L \; p^3, \qquad |p/L| \ll 1 , 
  \label{eq:es_fit}
\end{equation}
where $\Delta'_L \ll \Delta_L$, as \cite{EislerPeschel.JSTAT.2013}.  The label
$p$ is chosen now as a half-odd integer $p= \pm 1/2, \pm 3/2, \dots, \pm
(L-1)/2$ when $L$ is even and as an integer $p= 0, \pm 1, \dots, \pm (L -1)/2$
for $L$ odd.  The EEs given by \eqref{eq:es_fit} correspond to the ones where
$\nu_p \simeq 1/2$ which therefore contribute the most to the entanglement
entropy $S_L$.  In fact, making the approximation $\epsilon_p \approx
\Delta_L\; p$, we can compute $S_L$ in the limit $L \gg 1$,
\begin{equation}
  S_L  =   \sum_{p} \left[  
    \frac{ \log( 1+ e^{\epsilon_p})}{ 1 + e^{\epsilon_p}} + 
    \frac{ \log( 1+ e^{-\epsilon_p})}{ 1 + e^{-\epsilon_p}} \right]
  \approx
  2 \int_{-\infty}^\infty dx\; 
  \frac{\log(1+\exp(\Delta_L  x))}{1+\exp(\Delta_L  x)} 
  =  \frac{\pi^2}{ 3 \, \Delta_L}.
\label{eq:approxint}
\end{equation}

This equation is rather interesting since it relates $S_L$ to the inverse of
the entanglement spacing $\Delta_L$ and connects with previous results in the
literature \cite{PeschelTruong.87, CardyPeschel.NPB.88, Okunishi.JPSJ.05,
  LeporiDeChiaraSanpera.13}.  First of all, in the critical case, that is
$z=0$, where $S_L \approx 1/6 \log L$, it implies that $\Delta_L \propto
1/\log L$, as shown in \cite{Peschel.JSTAT.04}.  This result has wider
implications that lead to the understanding of the ES as the energy spectrum
of a boundary CFT on a strip of effective width $\propto \log L$
\cite{Lauchli.13}.  The computation in equation \eqref{eq:approxint} is
similar to the one of reference \cite{Cardy_Calabrese.JSTAT.04} for the non
critical Ising and XXZ models, which leads to the equation $S_L = c/6 \log
\xi$ where $\xi$ is the correlation length and is proportional to the inverse
of the level spacing of the spectrum of the corner transfer matrix Hamiltonian
on these models.

\begin{figure}
  \centering
  \includegraphics[width=100mm]{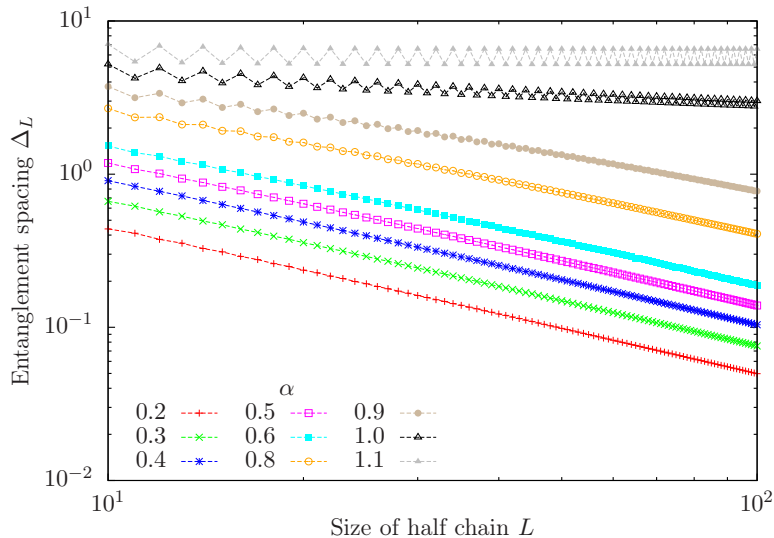}
  \caption{Entanglement spacing $\Delta_L$ as a function of $L$ for different
    values of $\alpha$.  Notice the behavior $\propto 1/\log L$ for $\alpha
    =1$ and $\propto 1/L$ for $\alpha < 1$ and $L$ large.  The case
    corresponding to $\alpha=1.1$ shows a qualitatively different behavior.}
    \label{fig:esclean}
\end{figure}

The dependence of the entanglement spacing $\Delta_L$ on the system size $L$
has a different behavior for $\alpha=1$ and $\alpha<1$.  Figure
\ref{fig:esclean} shows some $\Delta_L$ curves, for different values of
$\alpha$, in scale $\log L$.  As soon as $\alpha<1$ and large enough $L$ we
obtain a trend towards a power-law decay, which, for large $L$, converges to
$\Delta_L \approx 1/L$.  Combining this with \eqref{eq:approxint}, yields the
volume law for the entanglement entropy: $S(L)\approx 1/\epsilon \approx L$,
as expected.

\begin{figure}
  \includegraphics[width=80mm]{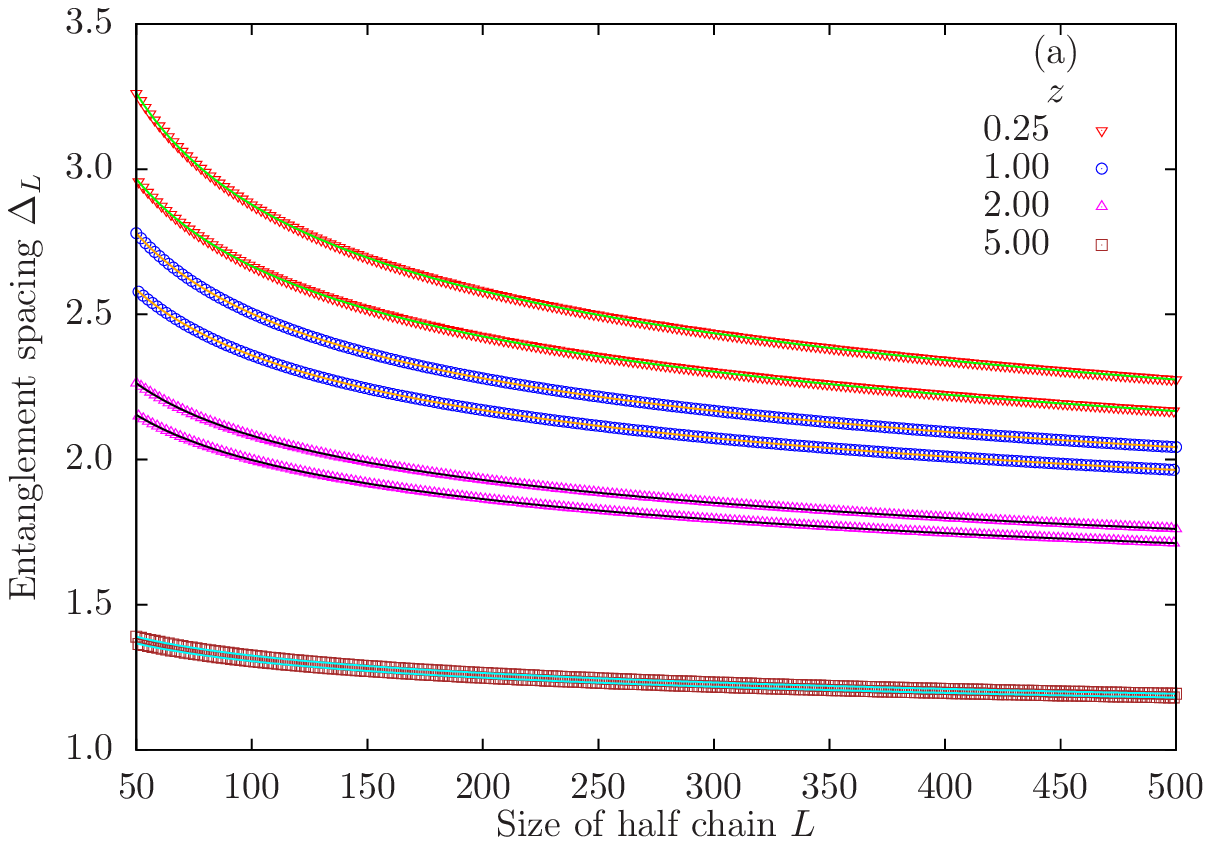}\\\vspace{5mm}
  \begin{minipage}[c]{40mm}
    \includegraphics[width=40mm]{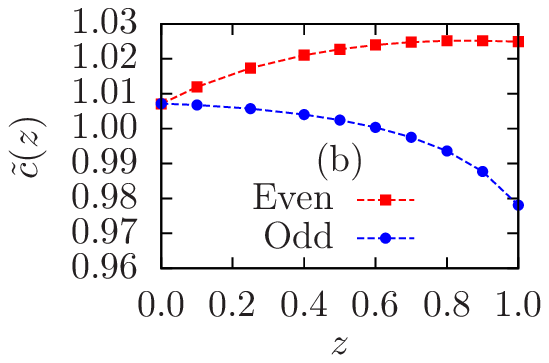}
  \end{minipage}\hspace{2mm}%
  \begin{minipage}[c]{40mm}
    \includegraphics[width=40mm]{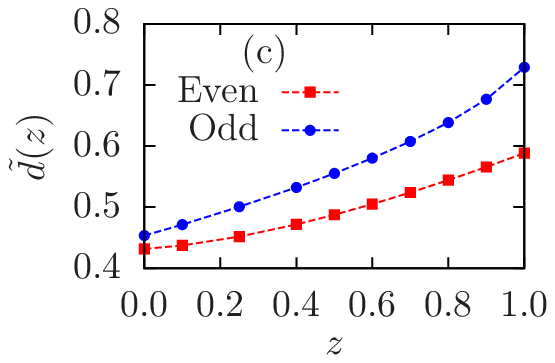}
  \end{minipage}\hspace{2mm}%
  \begin{minipage}[c]{40mm}
    \includegraphics[width=40mm]{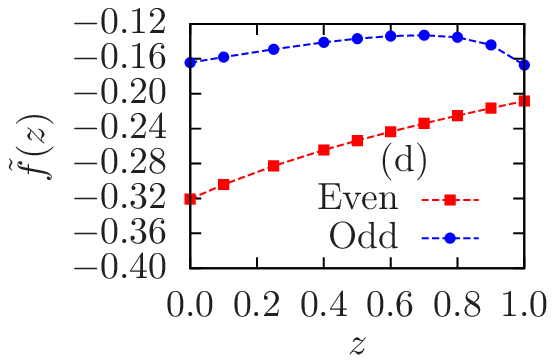}
  \end{minipage}\hspace{2mm}%
  \begin{minipage}[c]{40mm}
    \includegraphics[width=40mm]{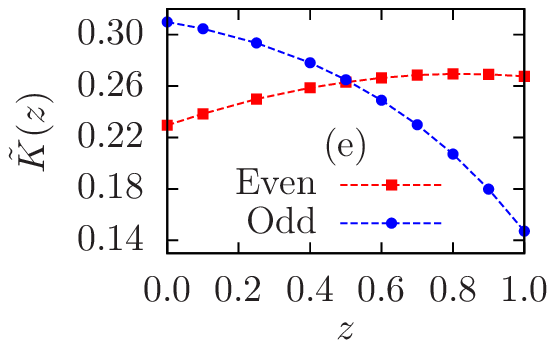}
  \end{minipage}
  \caption{(a) Entanglement spacing $\Delta_L$ as a function of the system
    half size $L=50, \dots, 500$.  For each value of $z$ the top (bottom)
    curves correspond to $L$ even (odd).  Solid lines are fits to
    \eqref{eq:fitgap} with $\chi^2 \sim \E{-12}$.  (b)-(e) Functions
    $\tilde{c}(z)$, $\tilde{d}(z)$, $\tilde{f}(z)$ and $\tilde{K}(z)$, in the
    interval $z \in [0, 1]$ for $L$ even (odd).}
  \label{fig:sl2}
\end{figure}

Based on equations \eqref{eq:sl} and \eqref{eq:approxint} we are led to the
following ansatz for the entanglement spacing
\begin{equation}
  \Delta_L \approx \frac{ \pi^2/3}
  {\frac{1}{6} \tilde{c}(z) \log L + \tilde{d}(z) + \tilde{f}(z) L^{-
      \tilde{K}(z)}},
  \label{eq:fitgap}
\end{equation}
where the functions $\tilde{c}(z)$, $\tilde{d}(z)$, $\tilde{f}(z)$ and
$\tilde{K}(z)$ depend on the parity of $L$.  This formula is extremely
accurate with a $\chi^2$ of order $\E{-12}$ in the range $z \in[0,1]$.  Figure
\ref{fig:sl2}(a) plots the values of $\Delta_L$ as a function of $L$ for
different values of $z$.  Notice that the parity oscillations of $L$ are
reminiscent to those of $S_L$.  The functions $\tilde{c}(z)$, $\tilde{d}(z)$
and $\tilde{f}(z)$ behave in a similar (though not identical) way to their
pairs $c(z)$, $d(z)$ and $f(z)$ in the interval $z \in [0,1]$, especially for
the $L$ even values.  For larger $z$ those fits lose quality.  Notice that
$K(z)$ is not $1$, but close to $0.25$.

\begin{figure}
  \centering
  \includegraphics[width=100mm]{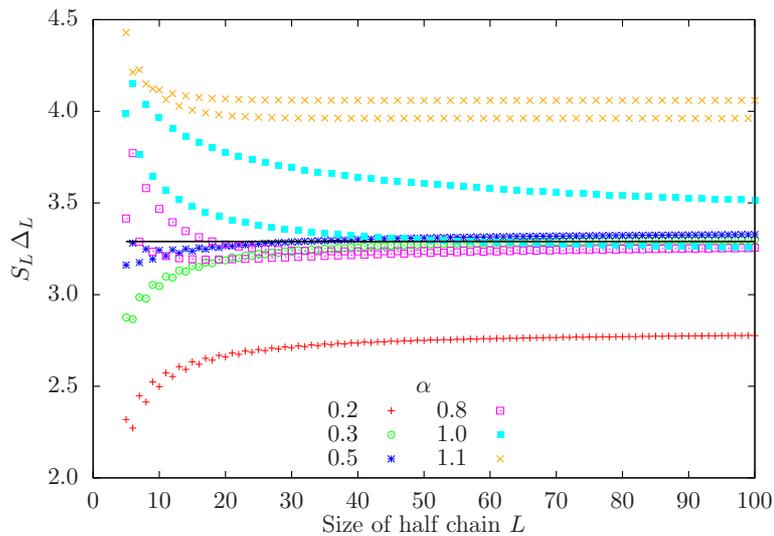} 
  \caption{Plot of the product $S_L \Delta_L$ to illustrate equation
    \eqref{eq:approxint}.  The black straight line is the constant $\pi^2/3$.}
  \label{fig:es_alpha}
\end{figure}

Finally, in order to verify equation \eqref{eq:approxint} we plot in figure
\ref{fig:es_alpha} the product $S_L \Delta_L$, which shows that for $\alpha
\leq 1$ the curves approach the constant $\pi^2/3$ for large values of $L$,
but not for $\alpha = 1.1$, which corresponds to a model with different
qualitative behavior.

\subsection{The spin 1/2 deformed  Heisenberg model}
\label{seq:heisenberg}

The deformed XX model can be immediately generalized to any 1D Hamiltonian
with OBC; $H = \sum_{i=1}^{2L -1} h_{i,i+1}$, whose exponential deformation is
given by
\begin{equation}
  H_L(\alpha) = J_0(\alpha)  h_{1, -1} +  \sum_{i=1}^{L-1} J_i(\alpha)\;  (
  h_{i,i+1} + h_{-i, - (i+1)}  ) ,
  \label{eq:generic.rainbow}
\end{equation}
where $J_i(\alpha)$ are defined in equation \eqref{eq:rb.couplings}.  We shall
next consider the exponential deformation of the Heisenberg Hamiltonian
defined by $h_{i,i+1}=\vec S_i \cdot \vec S_{i+1}$, where $\vec S_i$ are the
spin 1/2 matrices.  The deformed Hamiltonian easily follows from equation
\eqref{eq:generic.rainbow}.  The Dasgupta-Ma RG equation for the couplings is
given by \cite{Refael_Moore.JPA.09}
\begin{equation}
  \tilde J_i = \frac{J_{i-1} J_{i+1}}{2J_i}, 
  \label{eq:dasgupta-ma_heisenberg}
\end{equation}
which differs from equation \eqref{eq:dasgupta-ma} by a factor of $2$ in the
denominator.  In the limit $\alpha \rightarrow 0$, one obtains again the
rainbow state made of valence bonds across the middle of the chain.

\begin{figure}
  \centering
  \begin{minipage}[c]{80mm}
    \includegraphics[width=80mm]{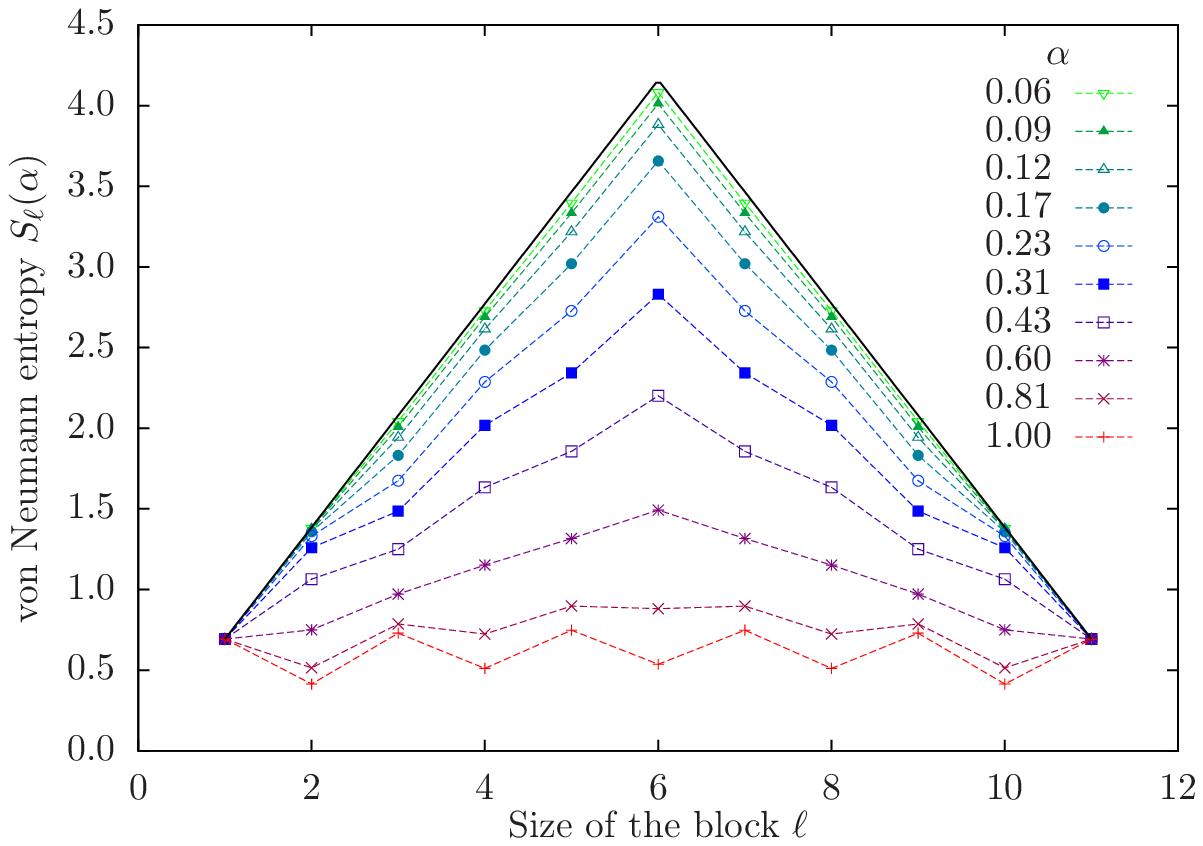}
  \end{minipage}%
  \hspace{10mm}%
  \begin{minipage}[c]{80mm}
    \includegraphics[width=80mm]{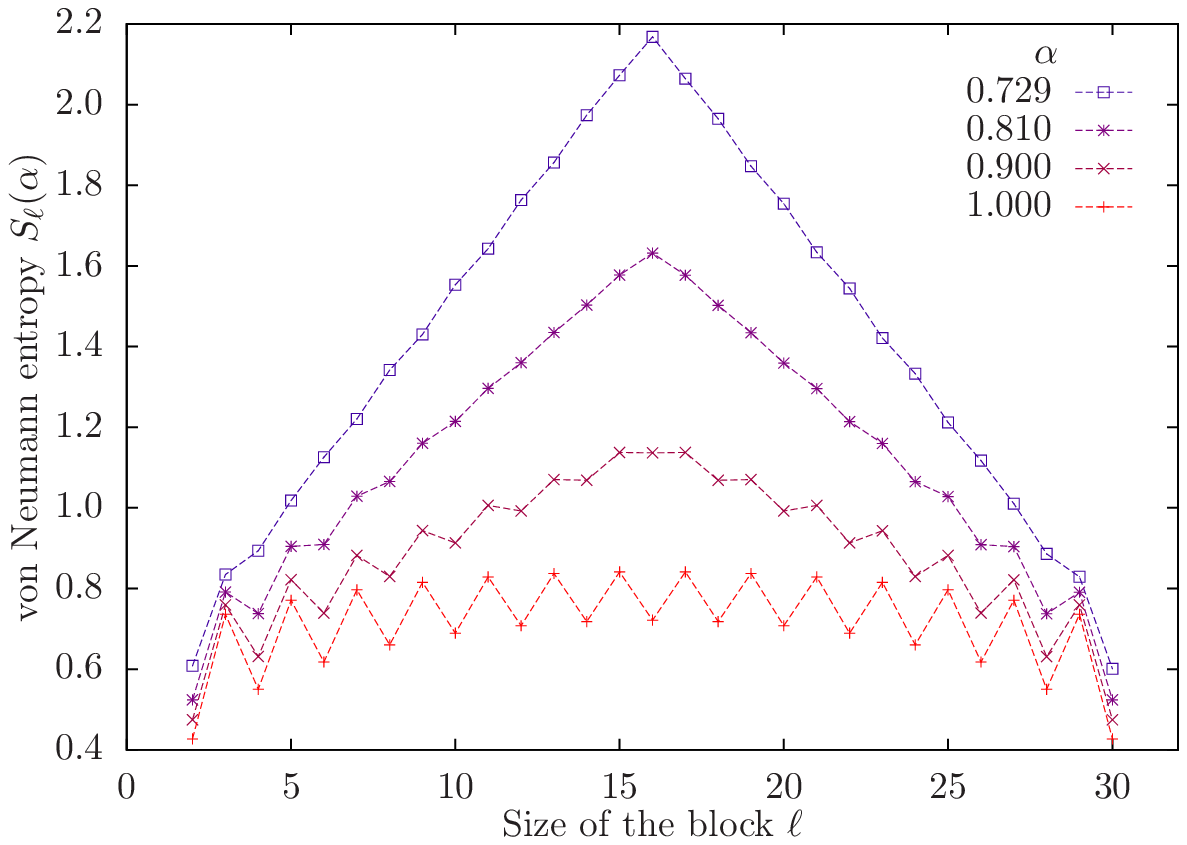}
  \end{minipage}
  \caption{Left: von Neumann entropy of the deformed Heisenberg system and
    $L=6$ ($12$ sites), where all the values of $\alpha$ are explored.  Right:
    system with $L=16$ ($32$ sites) studied with the DMRG method.  The fast
    increase of entanglement of the half-chain limits the range of $\alpha$
    where the method can be applied.}
  \label{fig:heisenberg.dmrg}
\end{figure}

The numerical study of the uniform to rainbow transition is more involved than
in the free fermionic case, because the GS cannot be obtained via single-body
procedures.  For very small system sizes, we have used exact diagonalization
of the many-body Hamiltonian, while for larger sizes we have employed the
density matrix renormalization group (DMRG).  The problem with the latter is
that we cannot reach very low values of $\alpha$, since the entanglement
entropy $S_L$ grows linearly with the system size and the number of retained
states grows exponentially with $S_L$.  Figure \ref{fig:heisenberg.dmrg}
summarizes our results: the left panel shows the exact von Neumann entropy as
a function of the block size $S_\ell(\alpha)$ for $L=6$ ($12$ sites).  Notice
the black line, which marks the Dasgupta-Ma RG limit.  The right panel shows
the same function for a system with $L=16$ ($32$ sites), but where $\alpha$
varies in the range $[0.7,1]$.  In both cases we can see the development of
the tent shape, which is the hallmark of the volume-law.

\section{Conclusions}
\label{sec:conclusions}

We have analyzed a deformation of critical local 1D Hamiltonians, which
interpolates between a log law and a volume law for the entanglement entropy.
The couplings between neighboring sites decay exponentially, with a factor
$\alpha^2$, as we move away from the middle point.  The value $\alpha=1$
corresponds to the uniform model, described by CFT and in the $\alpha\to 0^+$
limit the GS becomes a rainbow state, in which sites symmetrically placed with
respect to the center are maximally entangled.  There is a smooth crossover
between the uniform and the rainbow states that we have studied in detail for
the XX model (free spinless fermions model) and shown to be qualitatively
equivalent in the Heisenberg model.

In the XX model, the von Neumann entropy of any block, at not too small values
of $\alpha$, can be approximated by a combination of the CFT law plus a volume
law.  We have also found a scaling variable $z$ that depends on the size of
the chain and the magnitude of the deformation $\alpha$ in terms of which the
half-chain entanglement entropy is a renormalized version of the CFT formulas,
with coefficients depending on $z$.  We have discussed this result in
connection with the $c-$ and $g-$theorems.

The analysis of the ES shows very interesting connections between the
conformal growth, $S\sim \log L$ and the volumetric growth, $S \sim L$.
Indeed, the spectrum is approximately equally spaced, with an entanglement
spacing $\Delta_L$ that decays with the system size as $1/\log L$ at the
conformal point and as $1/L$ for $\alpha<1$.  We have also found that the
entanglement entropy is approximately proportional to the inverse of the
entanglement spacing, in wide regions of the parameter space, which
generalizes previous known results in critical and massive systems.

In summary, we have shown that an exponential deformation of the XX and
Heisenberg models offers the possibility to analyze the departure from the log
law of the entanglement entropy in CFT towards a volume law that is related to
the valence bond picture of these models.  It would be worth to study other
critical models to verify the generality of these results, as well as non
critical models that will exhibit a crossover from the area to the volume law.
Finally, it will be very interesting to construct the field theory underlying
the exponential perturbation of CFT's that will give an explanation of the
scaling behavior obtained numerically in this work \cite{laguna}.

\section{Acknowledgments}
We would like to acknowledge J~I Latorre, A Trombettoni, F~C Alcaraz, G
Mart\'{\i}nez and J Eisert.  We also acknowledge financial support from the
Spanish government from grant FIS2012-33642 and the Spanish MINECO Centro de
Excelencia Severo Ochoa Programme under grant SEV-2012-0249.  J~R-L
acknowledges support from grant FIS2012-38866-C05-1.  G~R acknowledges the
support from grant FIS2009-11654.

\bibliography{rainbow}
\bibliographystyle{apsrev4-1} 

\end{document}